\documentstyle[12pt]{article}

\def\noi{\noindent}

\def\nqq{\hspace{-2em}}

\def\barr{\left(\begin{array}}
\def\earr{\end{array}\right)}
\def\beq#1{\begin{equation}\label{#1}}
\def\eeq{\end{equation}}
\def\ber#1{\begin{eqnarray}\label{#1} \nqq}
\def\eer{\end{eqnarray}}
\def\eern{\nonumber \end{eqnarray}}
\def\nn{\nonumber\\ \nqq}
\def\mm{\\ \nqq}
\newcommand{\bear}[1]{\begin{eqnarray}\label{#1}}
\newcommand{\ear}{\end{eqnarray}}

\catcode`\@=11 \@addtoreset{equation}{section}\catcode`\@=12

\newcommand{\R}{\mbox{\bf R}}
\newcommand{\C}{\mbox{\bf C}}

\newcommand{\N}{\mbox{\bf N}}

\newcommand{\sign}{\mathop{\rm sign}\nolimits}

\newcommand{\sh}{\mathop{\rm sh}\nolimits}
\newcommand{\ch}{\mathop{\rm ch}\nolimits}

\newcommand{\eps}{\varepsilon}
\newcommand{\tri}{\triangle}

\newcommand{\e}[1]{\mathop{\rm e}\nolimits^{#1}}

\newcommand{\p}{\partial}

\newcommand{\fnm}{\footnotemark}
\newcommand{\fnt}{\footnotetext}

\begin{document}

\begin{center}
\large\bf
SOLUTIONS  WITH INTERSECTING P-BRANES
RELATED TO TODA CHAINS
\\[15pt]
\normalsize\bf V.D. Ivashchuk\fnm[1]\fnt[1]{ivas@rgs.phys.msu.su},
and  S.-W. Kim
\fnm[2]\fnt[2]{sungwon@mm.ewha.ac.kr} \\[10pt]

\it Center for Gravitation and Fundamental Metrology,
VNIIMS, 3/1 M. Ulyanovoy Str.,
Moscow 117313, Russia${}^{1
}$

\it Department of Science Education
and Basic Science Research Institute, Ewha Womans University, Seoul 120-750,
Korea${}^2$
\end{center}

\vspace{15pt}

\small\noi

\begin{abstract}

Solutions  in  multidimensional gravity
with $m$ $p$-branes  related to Toda-like systems (of general type)
are obtained. These solutions are defined on a product
of $n+1$ Ricci-flat spaces $M_0 \times M_1 \times  \ldots  \times M_n$
and are governed by one harmonic function on $M_0$. The solutions are
defined up to the solutions of Laplace and Toda-type equations
and correspond to null-geodesics of the
(sigma-model) target-space metric.
Special solutions  relating to $A_m$ Toda chains  (e.g. with $m =1,2$)
are considered.

\end{abstract}

\vspace{10cm}

\pagebreak

\normalsize

\section{Introduction}

 At present there exists a special interest to the so-called $M$ theory
(see, for example, \cite{M-th1}-\cite{M-th2}).
This theory is ``supermembrane'' analogue of
superstring models \cite{GSW} in $D=11$. The low-energy limit of
$M$-theory after a dimensional reduction leads to models governed by a
Lagrangian containing metric, fields of  forms and scalar fields.
These models contain a large variety of so-called
$p$-brane solutions (see \cite{St}-\cite{GM2} and references
therein).

In \cite{IMC} it was shown that after dimensional reduction on the
manifold $M_0\times M_1\times\dots\times M_n$ and when the composite
$p$-brane ansatz for fields of forms
is considered the problem is reduced to the gravitating
self-interacting $\sigma$-model with certain constraints imposed. (For
electric $p$-branes see also \cite{IM0,IM,IMR}.) This representation
may be considered as a powerful tool for obtaining different solutions
with intersecting $p$-branes (analogues of membranes). In
\cite{IMC,IMR,IMBl,IKM,GrI}
the Majumdar-Papapetrou type solutions (see \cite{MP})
were obtained (for non-composite
case see \cite{IM0,IM}). These solutions correspond to Ricci-flat
$(M_i,g^i)$, ($g^i$ is metric on $M_i$) $i=1,\dots,n$, and were also
generalized to the case of Einstein internal spaces \cite{IMC}.  Earlier
some special classes of these solutions were considered in
\cite{Ts1,PT,GKT,AR,AEH,AIR}. The obtained solutions take place, when
certain (block-)orthogonality relations (on couplings parameters,
dimensions of "branes", total dimension) are imposed. In this situation a
class of cosmological and spherically-symmetric solutions was obtained
\cite{IMJ,IMJ2}. Special cases were also considered in
\cite{LPX,BGIM,GrIM,BKR}.  The solutions with the horizon were considered
in details in \cite{CT,AIV,Oh,IMJ,BIM}.

In models under consideration  there exists a large variety of
Toda-chain solutions,  when certain intersection rules are satisfied
\cite{IMJ}.  Cosmological and spherically symmetric solutions with
$p$-branes and $n$ internal spaces  related to $A_m$ Toda chains
were previously  considered in   \cite{LPX,LMPX} and   \cite{GM1,GM2}.

It is well known
that geodesics of the target space equipped with some
harmonic function on a three-dimensional space generate a solution
to the $\sigma$-model equations \cite{NK,KSMH}.
(It was  observed in
\cite{GC} that  null geodesics of the target space of stationary
five-dimensional Kaluza-Klein theory may be used to generate multisoliton
solutions similar to the Israel-Wilson-Perj\`es solutions
of Einstein-Maxwell theory.)
Here we apply this null-geodesic method  to our sigma-model and
obtain a  new class of solutions in
multidimensional gravity with  $p$-branes
governed by one harmonic function $H$. The solutions
>from this class correspond to
null-geodesics of the target-space metric and are defined
by some functions $f_s(H)= \exp(-q^s(H))$
with $q^s(u)$  being  solutions to Toda-type equations.


\section{The model}

We consider a  model governed by
the action \cite{IMC}
\ber{1.1}
S=\int d^Dz\sqrt{|g|}\biggl\{R[g]-h_{\alpha\beta}g^{MN}\p_M\varphi^\alpha
\p_N\varphi^\beta-\sum_{a\in\tri}\frac{\theta_a}{n_a!}
\exp[2\lambda_a(\varphi)](F^a)^2\biggr\}
\eer
where $g=g_{MN}dz^M\otimes dz^N$ is a metric,
$\varphi=(\varphi^\alpha)\in\R^l$ is a vector of scalar fields,
$(h_{\alpha\beta})$ is a  constant symmetric
non-degenerate $l\times l$ matrix $(l\in \N)$,
$\theta_a=\pm1$, $F^a=dA^a$ is a $n_a$-form ($n_a\ge1$), $\lambda_a$ is a
1-form on $\R^l$: $\lambda_a(\varphi)=\lambda_{\alpha a}\varphi^\alpha$,
$a\in\tri$, $\alpha=1,\dots,l$. Here $\tri$ is some finite set.

We consider a manifold
\ber{1.2}
M=M_0\times M_1\times\dots\times M_n,
\eer
with a metric
\ber{1.3}
g=\e{2\gamma(x)}g^0+\sum_{i=1}^n\e{2\phi^i(x)}g^i
\eer
where $g^0=g_{\mu\nu}^0(x)dx^\mu\otimes dx^\nu$ is a metric on the
manifold $M_0$, and $g^i=g_{m_in_i}^i(y_i)dy_i^{m_i}\otimes dy_i^{n_i}$
is an Einstein metric on $M_{i}$ satisfying the equation
\beq{1.4}
R_{m_{i}n_{i}}[g^i ] = \xi_{i} g^i_{m_{i}n_{i}},
\eeq
$m_{i},n_{i}=1, \ldots, d_{i}$; $\xi_{i}= {\rm const}$,
$i=1,\ldots,n$.
(Here we identify notations  for $g^{i}$  and  $\hat{g}^{i}$, where
$\hat{g}^{i} = p_{i}^{*} g^{i}$ is the
pullback of the metric $g^{i}$  to the manifold  $M$ by the
canonical projection: $p_{i} : M \rightarrow  M_{i}$, $i = 0,
\ldots, n$. An analogous agreement will be also kept for volume forms etc.)

Any manifold $M_\nu$ is supposed to be
oriented and connected and $d_\nu\equiv\dim M_\nu$,
$\nu=0,\dots,n$. Let
\ber{1.5}
\tau_i \equiv\sqrt{|g^i(y_i)|}dy_i^1\wedge\dots\wedge dy_i^{d_i}, \quad
\eps(i)\equiv\sign(\det(g_{m_in_i}^i))=\pm1
\eer
denote the volume $d_i$-form and signature parameter respectively,
$i=1,\dots,n$. Let $\Omega=\Omega_n$ be a set of all subsets of
$\{1,\dots,n\}$, $|\Omega|=2^n$. For any $I=\{i_1,\dots,i_k\}\in\Omega$,
$i_1<\dots<i_k$, we denote
\ber{1.6}
\tau(I)\equiv\tau_{i_1}\wedge\dots\wedge\tau_{i_k},
\quad d(I)\equiv\sum_{i\in I}d_i,
\quad \eps(I) \equiv \prod_{i \in I} \eps(i).
\eer
We also put $\tau(\emptyset)= \eps(\emptyset)=
1$ and $d(\emptyset)=0$.

For fields of forms we consider the following composite electromagnetic
ansatz
\ber{1.7}
F^a=\sum_{I\in\Omega_{a,e}}{\cal F}^{(a,e,I)}+
\sum_{J\in\Omega_{a,m}}{\cal F}^{(a,m,J)}
\eer
where
\ber{1.8}
{\cal F}^{(a,e,I)}=d\Phi^{(a,e,I)}\wedge\tau(I), \mm
\label{1.9}
{\cal F}^{(a,m,J)}=\e{-2\lambda_a(\varphi)}*(d\Phi^{(a,m,J)}
\wedge\tau(J))
\eer
are elementary forms of electric and magnetic types respectively,
$a\in\tri$, $I\in\Omega_{a,e}$, $J\in\Omega_{a,m}$ and
$\Omega_{a,e}$, $\Omega_{a,m}$ are non-empty subsets of $\Omega$. In
(\ref{1.9}) $*=*[g]$ is the Hodge operator on $(M,g)$. For scalar
functions we put
\ber{1.10}
\varphi^\alpha=\varphi^\alpha(x), \quad
\Phi^s=\Phi^s(x),
\eer
$s\in S$, $x \in M_0$.

Here and below
\ber{1.11}
S=S_e \cup S_m, \quad
S_v=\bigcup_{a\in\tri}\{a\}\times\{v\}\times\Omega_{a,v},
\eer
$v=e,m$. The set $S$ consists of elements $s=(a_s,v_s,I_s)$,
where $a_s \in \tri$, $v_s = e, m$ and $I_s \in \Omega_{a_s,v_s}$.

Due to (\ref{1.8}) and (\ref{1.9})
\ber{1.12}
d(I)=n_a-1, \quad d(J)=D-n_a-1,
\eer
for $I\in\Omega_{a,e}$, $J\in\Omega_{a,m}$.

\subsection{The sigma model}

Let $d_0 \neq 2$ and
\ber{1.13}
\gamma=\gamma_0(\phi) \equiv
\frac1{2-d_0}\sum_{j=1}^nd_j\phi^j,
\eer
i.e. the generalized harmonic gauge is used.

We impose the restriction on sets $\Omega_{a,v}$.
These restrictions guarantee the block-diagonal structure
of a stress-energy tensor (like for the metric) and the existence of
$\sigma$-model representation \cite{IMC}.

We denote $w_1\equiv\{i|i\in\{1,\dots,n\},\quad d_i=1\}$, and
$n_1=|w_1|$ (i.e. $n_1$ is the number of 1-dimensional spaces among
$M_i$, $i=1,\dots,n$).

{\bf Restriction 1.} Let 1a) $n_1\le1$ or 1b) $n_1\ge2$ and for
any $a\in\tri$, $v\in\{e,m\}$, $i,j\in w_1$, $i<j$, there are no
$I,J\in\Omega_{a,v}$ such that $i\in I$, $j\in J$ and $I\setminus\{i\}=
J\setminus\{j\}$.

{\bf Restriction 2} (only for $d_0=1,3$). Let 2a) $n_1=0$ or
2b) $n_1\ge1$ and for any $a\in\tri$, $i\in w_1$ there are no
$I\in\Omega_{a,m}$, $J\in\Omega_{a,e}$ such that $\bar I=\{i\}\sqcup J$
for $d_0 = 1$ and
$J=\{i\}\sqcup \bar I$ for $d_0 = 3$. Here and in what
follows
\ber{1.13a}
\bar I\equiv\{1,\ldots,n\}\setminus I.
\eer
These restrictions are  satisfied in the non-composite case
\cite{IM0,IM}:  $|\Omega_{a,v}| = 1$, (i.e when there are no two
$p$-branes with the same color index $a$, $a\in\tri$.) Restriction 1 and 2
forbid certain intersections of two $p$-branes with the same color index
for  $n_1 \geq 2$ and  $n_1 \geq 1$ respectively.

It was proved in \cite{IMC} that equations of motion for the model
(\ref{1.1}) and the Bianchi identities: $d{\cal F}^s=0$, $s\in S_m$, for
fields from (\ref{1.3})--(\ref{1.13}), when Restrictions 1 and 2 are
imposed, are equivalent to equations of motion for the $\sigma$-model
governed by the action
\ber{2.1}
S_\sigma=\int d^{d_0}x\sqrt{|g^0|} \{R[g^0]-\hat G_{AB}
g^{0\mu\nu}\p_\mu z^A\p_\nu z^B \nn
-\sum_{s\in S}\eps_s\e{-2U_A^s z^A}
g^{0 \mu \nu} \p_\mu \Phi^s\p_\nu\Phi^s - 2V \},
\eer
where $(z^A)=(\phi^i,\varphi^\alpha)$, the index set
$S$ is defined in (\ref{1.11}),
\beq{2.1a}
 V = {V}(\phi)  = -\frac{1}{2}\sum_{i =1}^{n} \xi_i d_i
 e^{-2 \phi^i + 2 {\gamma_0}(\phi)}
\eeq
is the potential,
\ber{1.14}
(\hat G_{AB})=\barr{cc}
G_{ij}& 0\\
0& h_{\alpha\beta}
\earr,
\eer
is the target space metric with
\ber{1.15}
G_{ij}= d_i \delta_{ij}+\frac{d_i d_j}{d_0-2},
\eer
\ber{1.17}
(U_A^s)=(d_i\delta_{iI_s},-\chi_s\lambda_{\alpha a_s}),
\eer
are vectors,
$s=(a_s,v_s,I_s)$, $\chi_e=+1$, $\chi_m=-1$;
\ber{i}
\delta_{iI}=\sum_{j\in I}\delta_{ij}
\eer
is the indicator of $i$ belonging
to $I$: $\delta_{iI}=1$ for $i\in I$ and $\delta_{iI}=0$ otherwise; and
\ber{1.18}
\eps_s=(-\eps[g])^{(1-\chi_s)/2}\eps(I_s) \theta_{a_s},
\eer
$s\in S$, $\eps[g]\equiv\sign\det(g_{MN})$. More explicitly
(\ref{1.18}) reads $\eps_s=\eps(I_s) \theta_{a_s}$ for
$v_s = e$ and $\eps_s=-\eps[g] \eps(I_s) \theta_{a_s}$, for
$v_s = m$.

\section{Exact solutions  with one harmonic function}

\subsection{Toda-like Lagrangian}

Action (\ref{2.1}) may be also written in the form
\beq{2.1n}
S_\sigma=\int d^{d_0}x\sqrt{|g^0|}
\{ R[g^0]-  {\cal G}_{\hat A\hat B}(X)
g^{0 \mu \nu} \p_\mu X^{\hat A} \p_\nu X^{\hat B}   - 2V \}
\eeq
where $X = (X^{\hat A})=(\phi^i,\varphi^\alpha,\Phi^s)\in{\bf
R}^{N}$, and minisupermetric
\beq{2.1nn}
{\cal G}=
{\cal G}_{\hat A\hat B}(X)dX^{\hat A}\otimes dX^{\hat B} \eeq
on minisuperspace
\beq{2.1m}
{\cal M}={\bf R}^{N}, \quad   N = n+l+|S|
\eeq
($|S|$ is the number of elements in $S$) is defined by the relation
\beq{2.3n}
({\cal G}_{\hat A\hat B}(X))=\left(\begin{array}{ccc}
G_{ij}&0&0\\[5pt]
0&h_{\alpha\beta}&0\\[5pt]
 0&0&\eps_s \exp(-2U^s(X))\delta_{ss'}
\end{array}\right).
\eeq

Here we consider exact solutions to field equations corresponding
to the action  (\ref{2.1n})
\bear{2.4}
R_{\mu\nu}[g^0]=
{\cal G}_{\hat A \hat B}(X) \p_{\mu} X^{\hat A} \p_\nu  X^{\hat B}
+ \frac{2V}{d_0-2}g_{\mu\nu}^0,    \\
\label{2.5}
\frac{1}{\sqrt{|g^0|}} \p_\mu [\sqrt{|g^0|}
{\cal G}_{\hat C \hat B}(X)g^{0\mu\nu }  \p_\nu  X^{\hat B}]
- \frac{1}{2} {\cal G}_{\hat A \hat B, \hat C}(X)
g^{0,\mu \nu} \p_{\mu} X^{\hat A} \p_\nu  X^{\hat B} = V_{,\hat C},
\ear
$s\in S$. Here  $\tri[g^0]$ is the Laplace-Beltrami operator
corresponding to $g^0$ and  $V_{,\hat C} = \p V / \p  X^{\hat C}$.

We put
\ber{2.6}
X^{\hat A}(x) =  F^{\hat A}(H(x)),
\eer
where $F: (u_{-}, u_{+}) \rightarrow \R^{N}$  is a smooth function,
$H: M_0 \rightarrow \R $ is a harmonic function on $M_0$, i.e.
\ber{2.7}
\tri[g^0]H=0,
\eer
satisfying  $u_{-} < H(x)  < u_{+}$ for all $x \in M_0$.

The substitution of (\ref{2.6}) into  eqs. (\ref{2.4}) and (\ref{2.5})
leads us  to the relations
\bear{2.9}
R_{\mu \nu}[g^0]=
{\cal G}_{\hat A \hat B}(F(u))
\dot F^{\hat A} \dot F^{\hat B}   \p_\mu H  \p_\nu H
+ \frac{2V}{d_0-2}g_{\mu\nu}^0,  \\
\label{2.10}
\left[ \frac{d}{du}
\left( {\cal G}_{\hat C \hat B}(F(u)) \dot F^{\hat B}   \right)
- \frac{1}{2} {\cal G}_{\hat A \hat B, \hat C}(F(u))
\dot F^{\hat A} \dot F^{\hat B} \right]
g^{0,\mu \nu}  \p_\mu H  \p_\nu H = V_{,\hat C},
\ear
where $u = H(x)$ and $\dot f = d f/du$ .

Let all spaces  $(M_{i},g^i)$  be Ricci-flat, i.e.
\beq{2.11}
R_{m_{i}n_{i}}[g^i ] = 0,
\eeq
$i=0,\ldots,n$. In this case the potential is zero : $V = 0$ and
the field equations (\ref{2.9}) and (\ref{2.10})
are satisfied identically
if $F = F(u)$ obey the Lagrange equations  for the Lagrangian
\beq{2.12}
L =  \frac{1}{2} {\cal G}_{\hat A\hat B}(F) \dot F^{\hat A}  \dot F^{\hat B}
\eeq
with the zero-energy constraint
\beq{2.13}
E =  \frac{1}{2} {\cal G}_{\hat A\hat B}(F) \dot F^{\hat A}
\dot F^{\hat B} = 0.
\eeq
This means that  $F: (u_{-}, u_{+}) \rightarrow  \R^N$
is a null-geodesic map for the minisupermetric   (\ref{2.1nn}).
Thus, we are led to the Lagrange system (\ref{2.12})
with the minisupermetric  ${\cal G}$   defined in (\ref{2.3n}).

The problem of integrability may be simplified if we integrate
the Lagrange equations corresponding to $\Phi^s$
(i.e. the Maxwell equations for $s\in S_e$ and
Bianchi identities for $s\in S_m$):
\bear{2.14}
\frac d{du}\left(\exp(-2U^s(z))\dot\Phi^s\right)=0
\Longleftrightarrow
\dot\Phi^s=Q_s \exp(2U^s(z)),
\ear
where $Q_s$ are constants, $s \in S$.
Here $(F^{\hat A})= (z^A, \Phi^s)$.
We put
\bear{2.15}
Q_s\ne0,
\ear
for all  $s \in S$.

For fixed $Q=(Q_s,s\in S)$ the Lagrange equations for the Lagrangian
(\ref{2.12})  corresponding to $(z^A)=(\phi^i,\varphi^\alpha)$,
when equations (\ref{2.14}) are substituted, are equivalent to the
Lagrange equations for the Lagrangian
\beq{2.16}
L_Q=\frac12\hat G_{AB}\dot z^A\dot z^B-V_Q,
\eeq
where
\beq{2.17}
V_Q=\frac12  \sum_{s\in S}  \eps_s Q_s^2 \exp[2U^s(z)],
\eeq
$(\hat G_{AB})$ are defined in  (\ref{1.14})
respectively. The zero-energy constraint (\ref{2.13}) reads
\beq{2.18}
E_Q= \frac12 \hat G_{AB}\dot z^A \dot z^B+ V_Q =0.
\eeq

\subsection{ Toda-type solutions}

Let us  define the scalar product as follows
\ber{2.2}
(U,U')=\hat G^{AB}U_AU'_B,
\eer
for $U,U'\in\R^{n+l}$, where $(\hat G^{AB})=(\hat G_{AB})^{-1}$.
The scalar products (\ref{2.2}) for vectors $U^s$  were calculated in
\cite{IMC}
\ber{2.19}
(U^s,U^{s'})=d(I_s\cap I_{s'})+\frac{d(I_s)d(I_{s'})}{2-D}+
\chi_s\chi_{s'}\lambda_{\alpha a_s}\lambda_{\beta a_{s'}} h^{\alpha\beta},
\eer
where $(h^{\alpha\beta})=(h_{\alpha\beta})^{-1}$; $s=(a_s,v_s,I_s)$ and
$s'=(a_{s'},v_{s'},I_{s'})$ belong to $S$.

Here we are interested in exact solutions for a special case
when the vectors $U^s \in {\bf R}^{n+l}$ satisfy the following conditions
\beq{2.20}
K_s =(U^s,U^s)\neq 0,
\eeq
for all $s\in S$, and a (``quasi-Cartan'') matrix
\beq{2.21}
(A_{ss'}) = \left( \frac{2(U^s,U^{s'})}{(U^{s'},U^{s'})} \right)
\eeq
is a non-degenerate one. Here  some ordering in $S$ is assumed.
It follows from (\ref{2.20}) and
the non-degeneracy of the matrix (\ref{2.21})
that the vectors $U^s, s \in S,$ are  linearly
independent. Hence, the number of the vectors
$U^s$ should not exceed the dimension of
${\bf R}^{n+ l}$, i.e.
\beq{2.23}
|S| \leq n+ l.
\eeq

>From (\ref{2.19})-(\ref{2.21})  we get
the following  intersection rules \cite{IMJ}
\beq{2.24}
d(I_s \cap I_{s'})= \frac{d(I_s)d(I_{s'})}{D-2}-
\chi_s\chi_{s'}\lambda_{a_s}\cdot\lambda_{a_{s'}} + \frac12 K_{s'} A_{s s'},
\eeq
$s \neq s'$.

The  exact solutions
to Lagrange equations corresponding to
(\ref{2.16}) with the potential (\ref{2.17}) could be readily obtained
using the relations from Appendix.
The solutions read
\beq{2.34}
z^A = \sum_{s \in S} \frac{U^{sA}}{(U^s,U^s)} q^s + c^A u + \bar{c}^A,
\eeq
where $q^s$ are solutions to Toda-type equations
\beq{2.35}
\ddot{q^s} = -  B_s \exp( \sum_{s' \in S} A_{s s'} q^{s'} ),
\eeq
with
\beq{2.36}
 B_s = 2 K_s A_s,  \quad  A_s =  \frac12  \eps_s Q_s^2,
\eeq
$s \in S$. These equations correspond to the Lagrangian
\beq{2.37}
L_{TL} = \frac{1}{4}  \sum_{s,s' \in S} K^{-1}_s  A_{s s'} \dot{q^s}
\dot{q^{s'}}
-  \sum_{s \in S} A_s  \exp( \sum_{s' \in S} A_{s s'} q^{s'} ).
\eeq

Vectors $c=(c^A)$ and $\bar c=(\bar c^A)$ satisfy the linear constraint
relations (see (\ref{A.18}) in  Appendix)
\bear{2.47}
U^s(c)= U^s_A c^A=
\sum_{i \in I_s}d_ic^i-\chi_s\lambda_{a_s\alpha}c^\alpha=0,
\\ \label{2.50}
U^s(\bar c)=  U^s_A \bar c^A=  \sum_{i\in I_s}d_i\bar c^i-
\chi_s\lambda_{a_s\alpha}\bar c^\alpha=0,
\ear
$s\in S$.

The contravariant components $U^{rA}= \hat G^{AB} U^r_B$ are \cite{IMC,IMJ}
\beq{2.43}
U^{si}= G^{ij}U_j^s= \delta_{iI_s}-\frac{d(I_s)}{D-2}, \quad
U^{s\alpha}= - \chi_s \lambda_{a_s}^\alpha,
\eeq
Here (as in \cite{IMZ})
\beq{2.44}
G^{ij}=\frac{\delta^{ij}}{d_i}+\frac1{2-D},
\eeq
$i,j=1,\dots,n$, are the components of the matrix inverse to
$(G_{ij})$ from (\ref{1.15}).

Using (\ref{2.34})  and   (\ref{2.43}) we obtain
\beq{2.45}
\phi^i=  \sum_{s\in S} h_s \left(\delta_{iI_s}-\frac{d(I_s)}{D-2}\right)
 q^s + c^i u+  \bar c^i,
\eeq
and
\beq{2.46}
\varphi^\alpha=  - \sum_{s\in S} h_s \chi_s\lambda_{a_s}^\alpha q^s
+c^\alpha u+\bar c^\alpha,
\eeq
$\alpha=1,\dots,l$, and  $i =1,\dots,n$.
For $\gamma_0$ from (\ref{1.13}) we get
\beq{2.51}
\gamma_0(\phi) = - \sum_{s\in S}\frac{d(I_s)}{D-2} h_s q^s +
c^0u+ \bar c^0,
\eeq
where we denote
\beq{2.52}
h_s = K_s^{-1}
\eeq
and
\beq{2.52a}
c^0 = \frac1{2-d_0}\sum_{j=1}^n d_j c^j,
\quad  \bar  c^0 = \frac1{2-d_0}\sum_{j=1}^n d_j \bar c^j.
\eeq

The zero-energy constraint reads (see Appendix)
\bear{2.53}
2E= 2E_{TL} + \hat G_{AB}c^Ac^B =   \\ \nonumber
2E_{TL} + h_{\alpha\beta}c^\alpha c^\beta+ \sum_{i=1}^n d_i(c^i)^2+
\frac1{d_0-2}\left(\sum_{i=1}^nd_ic^i\right)^2 = 0.
\ear
where
\beq{2.54a}
E_{TL} = \frac{1}{4}  \sum_{s,s' \in S} K^{-1}_s
A_{s s'} \dot{q^s} \dot{q^{s'}}
  + \sum_{s \in S} A_s  \exp( \sum_{s' \in S} A_{s s'} q^{s'} ),
\eeq
is an integration constant (energy) for the solutions from
(\ref{2.35}).

>From the relation
\beq{2.56}
\exp(2 U^s(z)) = \prod_{s' \in S}  f_{s'}^{- A_{s s'}},
\eeq
following from  (\ref{2.21}), (\ref{2.34}),
(\ref{2.47}) and (\ref{2.50}) we get for
electric-type forms (\ref{1.8})
\beq{2.57}
{\cal F}^s= Q_s
\left( \prod_{s' \in S}  f_{s'}^{- A_{s s'}} \right) dH \wedge\tau(I_s),
\eeq
$s\in S_e$, and for magnetic-type forms (\ref{1.9})
\beq{2.58}
 {\cal F}^s=\exp[-2\lambda_a(\varphi)]
* \left[Q_s \left( \prod_{s' \in S} f_{s'}^{- A_{s s'}} \right)
 dH \wedge \tau(I_s)\right] =
\bar Q_s (*_0 d H) \wedge \tau(\bar I_s),
\eeq
$s\in S_m$, where  $\bar Q_s=Q_s\eps(I_s)\mu(I_s)$
and $\mu(I) =\pm1$ is defined by the relation
$\mu(I) dH \wedge \tau(I_0)=\tau(\bar I)\wedge dH \wedge\tau(I)$,
$I_0 = \{1, \dots, n \}$
(see eq. (2.26) in \cite{IMC}). Here $*_0 = *[g^0]$ is the Hodge operator
on $(M_0,g^0)$.

Relations for the metric and scalar fields follows from (\ref{2.45})-(\ref{2.51})
\bear{2.63}
g= \biggl(\prod_{s \in S} f_s^{2d(I_s)h_s/(D-2)}\biggr)
\biggl\{ \exp(2c^0 H+ 2\bar  c^0) g^0               \\ \nn
+ \sum_{i =1}^{n} \Bigl(\prod_{s\in S }
f_s^{- 2h_s \delta_{i I_s} }\Bigr)
\exp(2c^i H+ 2\bar  c^i) g^i \biggr\}, \\  \label{2.64}
\exp(\varphi^\alpha) =
\left( \prod_{s\in S} f_s^{h_s \chi_s\lambda_{a_s}^\alpha} \right)
\exp(c^\alpha H +\bar c^\alpha),
\ear
$\alpha=1,\dots,l$.
Here
\beq{2.65}
f_s = f_s(H) = \exp(- q^s(H)),
\eeq
where $q^s(u)$ is a solution to  Toda-like equations  (\ref{2.35})
and $H = H(x)$ ($x \in M_0$) is a harmonic function on $(M_0,g^0)$
(see (\ref{2.7})).

The solution is presented
by relations  (\ref{2.57})-(\ref{2.65})
with the functions  $q^s$  defined in
(\ref{2.35}) and the relations on the parameters of
solutions $c^A$, $\bar c^A$ $(A= i,\alpha,0)$,  $Q_s$ ,
$h_s$ imposed in  (\ref{2.47}),(\ref{2.50}), (\ref{2.52a}) and
(\ref{2.52}),  respectively.

This solution describes a set of charged (by forms) overlapping
$p$-branes ($p_s=d(I_s)-1$, $s \in S$) ``living'' on submanifolds
of $M_1 \times \dots \times M_n$.
The solution is valid if the dimensions of $p$-branes and dilatonic
coupling vector satisfy the relations (\ref{1.12}).

\section{Solutions corresponding to $A_m$  Toda  chain}

Here we  consider exact solutions to equations of motion
of a Toda-chain corresponding to the Lie algebra
 $A_{m}= sl(m+1, \C)$ \cite{T,And} ,
\beq{B.1}
\ddot q^s  = - B_s \exp\left( \sum_{s'=1}^{m} A_{s s'} q^{s'}  \right) ,
\eeq
where
\beq{B.1a}
\left(A_{ss'}\right)=
\left( \begin{array}{*{6}{c}}
2&-1&0&\ldots&0&0\\
-1&2&-1&\ldots&0&0\\
0&-1&2&\ldots&0&0\\
\multicolumn{6}{c}{\dotfill}\\
0&0&0&\ldots&2&-1\\
0&0&0&\ldots&-1&2
\end{array}
\right)\quad
\eeq
is the Cartan matrix  of the Lie algebra $A_{m}$ and $B_s > 0$,
$s,s' = 1, \ldots, m$.  Here we put
$S =  \{1, \ldots, m \}$.

The equations of motion   (\ref{B.1}) correspond
to the Lagrangian
\beq{B.2}
 L_T = \frac{1}{2} \sum_{s,s'=1}^{m} A_{ss'} \dot q^s  \dot q^{s'}  -
\sum_{s=1}^{m}  B_s \exp \left( \sum_{s'=1}^{m} A_{ss'} q^{s'}  \right).
\eeq
This Lagrangian may be obtained from the standard one \cite{T}
by separating a coordinate describing the  motion  of the center of mass.

Using the result of A. Anderson \cite{And}
we present the solution to eqs. (\ref{B.1}) in the following form
\beq{B.3}
C_s \exp(-q^s(u)) =
\sum_{r_1< \dots <r_s}^{m+1} v_{r_1}\cdots v_{r_s}
\Delta^2( w_{r_1}, \ldots, w_{r_s}) \exp[(w_{r_1}+\ldots +w_{r_s})u],
\eeq
$s = 1, \ldots, m$, where
\beq{B.4a}
\Delta( w_{r_1}, \ldots, w_{r_s})  =
\prod_{i<j}^{s} \left(w_{r_i}-w_{r_j}\right); \quad
\Delta(w_{r_1}) \equiv 1,
\eeq
denotes the  Vandermonde determinant.
The real constants $v_r$ and $w_r$, $r = 1, \ldots, m + 1$, obey the
relations
\beq{B.5}
\prod_{r=1}^{m+1} v_r= \Delta^{-2}(w_1,\ldots, w_{m+1}), \qquad
\sum_{r=1}^{m+1}w_r=0.
\eeq
In (\ref{B.3})
\beq{B.6}
C_s = \prod_{s'=1}^{m} B_{s'}^{-A^{s s'}},
\eeq
where
\ber{B.7}
A^{s s'}= \frac1{m+1}\min(s,s')[m+1-\max(s,s')],
\eer
$s, s' = 1, \ldots, m$,  are components of a matrix inverse to the Cartan one, i.e.
$(A^{s s'})=(A_{ss'})^{-1}$ (see Ch.7 in  \cite{FS}).
Here
\beq{B.8}
v_r \neq 0, \qquad w_r \neq w_{r'}, \quad r \neq r',
\eeq
$r, r' = 1, \ldots, m +1$.
We note that the solution with $B_s > 0$ may be obtained
>from the solution  with $B_s =1$ (see \cite{And}) by a certain shift $q^s
\mapsto q^s + \delta^s$.

The energy reads \cite{And}
\beq{B.9}
 E_T= \frac{1}{2} \sum_{s,s'=1}^{m} A_{s s'} \dot q^s  \dot q^{s'} +
\sum_{s=1}^{m}  B_s \exp \left( \sum_{s'=1}^{m} A_{ss'} q^{s'}  \right) =
\frac{1}{2}\sum_{r=1}^{m+1} w^2_r.
\eeq

If  $B_s > 0$, $s \in S$, then  all $w_r, v_r$ are real and, moreover, all
$v_r > 0$, $r =1, \ldots, m+1$.  In a general case $B_s \neq 0$, $s \in
S$, relations  (\ref{B.3})-(\ref{B.6}) also describe real solutions to
eqs. (\ref{B.1}) for suitably chosen complex
parameters $v_r$ and $w_r$. These parameters are either real or belong to
pairs of complex conjugate (non-equal) numbers, i.e., for example, $w_1 =
\bar w_2$, $v_1 = \bar v_2 $. When some of $B_s$ are negative, there are
also some special (degenerate) solutions to eqs. (\ref{B.1}) that are not
described by relations (\ref{B.3})-(\ref{B.6}), but may be obtained from
the latter by certain limits of parameters $w_i$ (see example in the next
section).

For the energy (\ref{2.54a}) we get
\beq{B.10}
E_{TL} = \frac{1}{2K} E_T=  \frac{h}{4} \sum_{r=1}^{m+1} w^2_r.
\eeq
Here
\beq{B.11}
K_s = K,    \qquad  h_s= h = K^{-1},
\eeq
$s \in S$.

Thus, in the $A_m$ Toda chain case  (\ref{B.1})
eqs.  (\ref{B.3})-(\ref{B.11}) should be substituted into
relations  (\ref{2.53}) and (\ref{2.57})-(\ref{2.65}) for the general
solution.

\subsection{Examples for $d_0 >2$}

Here we consider the case $d_0 > 2$. Let
matrix $(h_{\alpha \beta})$ be positively defined and
$K = K_s > 0$. Then from the energy constraint
(\ref{2.53})  we get
\beq{B.12a}
E_{TL} \leq 0 \Longrightarrow E_T \leq 0.
\eeq
In this case
\beq{B.13a}
c^\alpha = c^i = 0 \Longleftrightarrow E_{TL} = E_T = 0,
\eeq
$i =1, \ldots,n$, and $\alpha = 1, \ldots, l$.
When $(h_{\alpha \beta})$  is negatively defined
(as it takes place in $12$-dimensional theory from \cite{KKLP}) there exist
solutions with $E_{TL} > 0$.

\subsubsection{$A_1$-case.}

Here we consider the case of one ``brane'', i.e.  $S = \{ s \}$.
Solving the Liouville equation
\beq{B.14a}
\ddot q^s  = - B_s \exp(2 q^{s}),
\eeq
with $B_s = \eps_s K_s Q_s^2$, we
get
\beq{B.15a}
f_s(H) = |K_s|^{1/2} |Q_s| \hat f_s(H),
\eeq
where  $H = H(x) > 0$ and
\bear{B.12}
\hat f_s(H) =
\frac{1}{\sqrt{E_T}} \ch(\sqrt{E_T} H), \ \eps_s K_s > 0, \ E_T > 0;
 \\ \label{B.13}
\frac{1}{\sqrt{E_T}} \sh(\sqrt{E_T} H), \ \eps_s K_s <0, \ E_T > 0;
  \\ \label{B.14}
\frac{1}{\sqrt{-E_T}} \sin(\sqrt{-E_T} H), \ \eps_s K_s <0, \ E_T < 0;
   \\ \label{B.15}
                                      H, \ \eps_s K_s <0, \ E_T = 0.
\ear
Here
\beq{B.9b}
 E_T=  (\dot q^{s})^2 + B_s \exp(2 q^{s}).
\eeq

In a special case $E_T = 0$ this solution agrees with those from
\cite{IMC,IMBl} if
the following redefinition of the harmonic function is performed:
$H \mapsto \hat H$,
\beq{B.17}
\hat H = |K_s|^{1/2} |Q_s| H.
\eeq

\subsubsection{$A_2$-case.}

Now we consider the case  $m =2$
(for perfect fluid case see also \cite{GIM}).
Here we put $S = \{1,2 \}$.
The solution reads
\bear{C.1}
C_1 \exp(-q^1)  = v_1 \exp(w_1 u)  +   v_2 \exp(w_2 u) + v_3 \exp(w_3 u),
   \\ \label{C.2}
C_2 \exp(-q^2)  =  v_1 v_2 (w_1 - w_2)^2  \exp(-w_3 u)
  \\  \nonumber
 + v_2 v_3 (w_2 - w_3)^2  \exp(- w_1 u)
  + v_3 v_1 (w_3 - w_1)^2  \exp( -w_2 u),
\ear
where
\bear{C.3}
w_1 +  w_2 +  w_3 = 0, \\ \label{C.4}
v_1 v_2 v_3 =  (w_1 - w_2)^{-2} (w_2 - w_3)^{-2} (w_3 - w_1)^{-2} .
\ear
and
\beq{C.5}
 C_1 = (B_1^2 B_2)^{-1/3}, \qquad   C_2 = (B_2^2 B_1)^{-1/3}.
\eeq

Let $K > 0$. Then  $E_T \leq 0$ and hence some of $B_i$  should be negative.

Let  $B_1 < 0$ and  $B_2 < 0$. In the pseudo-Euclidean case, when
$\eps[g] = -1$
and all $\theta_a =1 $, this means that
$\eps_{s_i} = \eps(I_{s_i}) = -1$, $i =1,2$, (see (\ref{1.18})),
i.e. all $p$-branes should contain an odd number of  time submanifolds.

Let us consider  solutions  with a negative energy
\beq{C.5a}
  E_{T}  = \frac{1}{2} (w_1^2 + w_2^2 + w_3^2)  < 0.
\eeq
In this case two of parameters $w_i$  should be  complex.
Without loss of generality  we put
\bear{C.6}
w_1 = - 2\alpha,  \quad w_2 = \alpha + i \beta,
\quad    w_3 = \alpha - i \beta, \\ \label{C.7}
 v_2 = v e^{i \theta}, \quad   v_3 = v e^{-i \theta},
\ear
where parameters $ \beta \neq 0$, $\alpha$
and $v > 0$ are real and
\beq{C.8}
v_1 v^2 = - \frac{1}{4} \beta^{-2} (9 \alpha^2 + \beta^2)^{-2}.
\eeq

Then relations  (\ref{C.1}) and (\ref{C.2}) read
\bear{C.9}
|C_1| \exp(-q^1)  = - v_1 \exp( - 2 \alpha u)
+  2 v \exp( \alpha u)  \cos(\beta u + \theta),
   \\ \label{C.10}
|C_2| \exp(-q^2)  = 4 \beta^2 v^2 \exp( 2 \alpha u)  -
 2 v v_1  (9 \alpha^2 + \beta^2)
\exp(- \alpha u)  \cos(\beta u + \theta + 2 \varphi),
\ear
where $|C_1| = (B_1^2 |B_2|)^{-1/3}$,
$|C_2| = (B_2^2 |B_1|)^{-1/3}$, $|B_j| = Q_j^2$, $j =1,2$, and
\beq{C.9a}
3\alpha + i \beta =   (9 \alpha^2 + \beta^2)^{1/2} e^{i \varphi}.
\eeq

There exists also a degenerate solution with $E_T = 0$
\beq{C.10a}
|C_1| \exp(-q^1) =|C_2| \exp(-q^2) = \frac{1}{2} (u - u_0)^2,
\eeq
$u_0 = {\rm const}$, that may be obtained from  the solution
(\ref{C.9}),(\ref{C.10}) with $\alpha = 0$, $v_1 = -2v$, $v = 1/2
\beta^2$, $\theta = - \beta u_0$
\beq{C.11}
|C_1| \exp(-q^1) =|C_2| \exp(-q^2) =
\frac{2}{\beta^2} \sin^2[\beta(u - u_0) /2],
\eeq
in the limit $\beta \to 0$.

Let us consider an example of the  $A_2$-solution  in
$D = 11$ supergravity \cite{CJS}. We put $n = 3$,
$g^3=-dt\otimes dt$, $d_1= 2$, $d_2= 5$,
$d_0=3$ (metrics $g^0,g^1,g^2$ are Ricci-flat). The
$A_2$-solution, describing a dyon configuration with electric $2$-brane
and magnetic $5$-brane, corresponding to 4-form  $F$ and
intersecting in 1-dimensional time manifold  reads:
\bear{C.12}
g= c^{2/9} (\hat H^2 g^0- \hat H^{-2} dt \otimes dt)
+ c^{8/9}  g^1+  c^{-4/9} g^2,
\mm
\label{C.13}
F= c \nu_1 d\hat H^{-1}\wedge dt\wedge\tau_1+
c  \nu_2(*_0d \hat H)\wedge\tau_1,
\eer
where  $\hat H$ is the harmonic function on $(M_0,g^0)$
and $\nu_1^2 = \nu_2^2 =1$. This solution corresponds to the degenerate
solution (\ref{C.10a}).  Here the following  notations $\hat H = H (|Q_1|
|Q_2| )^{1/2}$ , $c = (|Q_2|/ |Q_1| )^{1/3}$ are adopted. For $c =1$ this
solution coincides with that of \cite{IMBl}.

\section{Cosmological-type solutions}

\subsection{Solutions with Ricci-flat spaces}

Let us consider a ``cosmological'' case:   $d_0 = 1$    and
\beq{3.1}
  M_0 = \R, \qquad  g^0 = w du\otimes du,
\eeq
where $w = \pm 1$. Since
$H(u)= u$ is a harmonic function on $(M_0,g^0)$  we get
for the metric and scalar fields from (\ref{2.63}), (\ref{2.64})
\bear{3.3}
g= \biggl(\prod_{s \in S} f_s^{2d(I_s)h_s/(D-2)}\biggr)
\biggl\{ \exp(2c^0 u+ 2\bar  c^0) w du \otimes du   \\ \nonumber
+ \sum_{i =1}^{n} \Bigl(\prod_{s\in S }
f_s^{- 2h_s \delta_{i I_s} }\Bigr)
\exp(2c^i u + 2\bar  c^i) g^i \biggr\},
\\  \label{3.4}
\exp(\varphi^\alpha) = \left( \prod_{s\in S}
f_s^{h_s \chi_s\lambda_{a_s}^\alpha} \right)
\exp(c^\alpha u +\bar c^\alpha),
\ear
$\alpha=1,\dots,l$, where   $f_s = f_s(u) = \exp(- q^s(u))$ and
$q^s(u)$ obey  Toda-like equations  (\ref{2.35}).

Relations (\ref{2.52a}) and  (\ref{2.53}) take the form
\bear{3.5}
c^0 = \sum_{j=1}^n d_j c^j,
\qquad  \bar  c^0 = \sum_{j=1}^n d_j \bar c^j, \\  \label{3.6}
2E= 2E_{TL} + h_{\alpha\beta}c^\alpha c^\beta+ \sum_{i=1}^n d_i(c^i)^2
- \left(\sum_{i=1}^nd_ic^i\right)^2 = 0,
\ear
with $E_{TL}$  from (\ref{2.54a}) and all other relations
(e.g. constraints (\ref{2.47}), (\ref{2.50})) and
relations  for forms (\ref{2.57}) and (\ref{2.58}) with $H = u$)
are unchanged.  In a special  $A_m$ Toda chain case this solution
was considered previously in \cite{GM2}.

\subsection{Solutions with one curved space}

The cosmological solution with Ricci-flat spaces
may be also  modified to the following case:
\beq{4c.2}
 \xi_1 \ne0, \quad \xi_2=\dots=\xi_n=0,
\eeq
i.e. one space is curved and others are Ricci-flat and
\beq{4c.3}
 1 \notin I_s,  \quad  \forall s  \in S,
\eeq
i.e. all ``brane'' submanifolds  do not  contain $M_1$.

Relation (\ref{4c.2}) modifies the potential $(\ref{2.17})$
by adding the term
\beq{4c.6}
\delta V=\frac12 w\xi_1 d_1  \exp(2U^1(z)),
\eeq
where ($d_1 > 1$)
\bear{4c.6a}
U^1(z)=U_A^1 z^A=-\phi^1+\gamma_0(\phi), \\ \label{4c.6b}
(U_A^1)=(-\delta_i^1+d_i,0).
\ear

For the scalar products we get \cite{IMJ}
\bear{4c.7}
(U^1,U^1)=\frac1{d_1}-1<0, \\ \label{4c.8}
(U^1,U^{s})=0
\ear
for all $s\in S$.

The solutions in the case under consideration
may be obtained   by a little modification
of the solution (\ref{2.34}) (see Appendix)
\beq{4c.9}
z^A(u)=-\frac{U^{1A}}{(U^1,U^1)}\ln |f_1(u-u_1)| -
\sum_{s\in S}\frac{U^{sA}}{(U^s,U^s)}\ln (f_s(u)) + c^A u +
\bar{c}^A,
\eeq
where
\bear{4c.10}
f_1(\tau) =R \sh(\sqrt{C_1}\tau), \ C_1>0, \ \xi_1 w>0;
\\ \label{4c.11}
R \sin(\sqrt{|C_1|}\tau), \ C_1<0, \  \xi_1 w>0;   \\ \label{4c.12}
R \ch(\sqrt{C_1}\tau),  \ C_1>0, \ \xi_1w <0; \\ \label{4c.13}
\left|\xi_1(d_1-1)\right|^{1/2} \tau, \ C_1=0,  \ \xi_1w>0,
\ear
$u_1$ and $C_1$ are constants and $R =  |\xi_1(d_1-1)/C_1|^{1/2}$.

Vectors $c=(c^A)$ and $\bar c=(\bar c^A)$ satisfy the linear constraints
(\ref{2.47}), (\ref{2.50}) and also additional constraints
\bear{4c.15}
U^1(c)= U^1_A c^A = -c^1+\sum_{j=1}^nd_jc^j=0, \\ \label{5.48n}
U^1(\bar c)= U^1_A \bar c^A =
-\bar c^1+\sum_{j=1}^nd_j\bar c^j=0.
\ear
The zero-energy constraint reads
\beq{4c.16}
E=E_1+ E_{TL}+ \frac12 \hat G_{AB} c^A c^B = 0,
\eeq
where $C_1=2E_1(U^1,U^1)$ or, equivalently,
\beq{4c.17}
C_1\frac{d_1}{d_1-1}= 2 E_{TL} +
h_{\alpha\beta}c^\alpha c^\beta+ \sum_{i=2}^nd_i(c^i)^2+
\frac1{d_1-1}\left(\sum_{i=2}^nd_ic^i\right)^2.
\eeq

>From  (\ref{4c.7}), (\ref{4c.9}) and
\beq{4c.18}
U^{1i}=-\frac{\delta_1^i}{d_1}, \quad U^{1\alpha}=0,
\eeq
we get a relation for the metric
\bear{4c.19}
g= \biggl(\prod_{s \in S} [f_s(u)]^{2 d(I_s) h_s/(D-2)} \biggr)
\biggl\{[f_1(u-u_1)]^{2d_1/(1-d_1)}\exp(2c^1u + 2 \bar c^1)\\ \nn
\times[w du \otimes du+ f_1^2(u-u_1)g^1] +
\sum_{i = 2}^{n} \Bigl(\prod_{s\in S}
[f_s(u)]^{- 2 h_s  \delta_{i I_s} } \Bigr)\exp(2c^i u+ 2 \bar c^i) g^i\biggr\}.
\ear
All other relations are unchanged. (Here $H(u)= u$
and $*_0 d H = w$ in (\ref{2.58}).)
This solution in a special case
of $A_m$ Toda chain was obtained earlier  in \cite{GM1}.

\section{Appendix: Solutions for Toda-like system }

Let
\bear{A.1}
L=\frac12<\dot x,\dot x>- \sum_{s \in S} A_s\exp(2<b_s,x>)
\ear
be a Lagrangian, defined on $V\times V$, where $V$ is
$n$-dimensional vector space over $\R$, $A_s\ne0$, $s \in S$;
$S \ne\emptyset$, and $<\cdot,\cdot>$ is
non-degenerate real-valued quadratic form on $V$.
Let
\bear{A.2}
K_s = <b_s,b_{s}> \neq 0,
\ear
for all $s \in S$.

Then, the Euler-Lagrange equations for the Lagrangian (\ref{A.1})
\beq{A.4}
\ddot{x} + \sum_{s \in S} 2 A_s b_s \exp(2<b_s,x>) =0,
\eeq
have the following  solutions
\beq{A.5}
x(u)=  \sum_{s \in S} \frac{q^s(u)  b_s}{<b_s,b_{s}>} +
\alpha u + \beta,
\eeq
where $\alpha,\beta\in V$,
\beq{A.6}
<\alpha,b_s>=<\beta,b_s>=0,
\eeq
$s \in S$, and functions $q^s(u)$
satisfy the Toda-like  equations
\beq{A.7}
\ddot{q^s} = -  2 A_s  K_s \exp( \sum_{s' \in S} A_{s s'} q^{s'} ),
\eeq
with
\beq{A.8}
 A_{s s'} = \frac{2 <b_s,b_{s'}>}{<b_{s'},b_{s'}>},
\eeq
$s, s'  \in S$. Let the matrix  $(A_{s s'})$ be a non-degenerate one.
In this case vectors $b_s$, $s \in S$, are linearly independent.
Then eqs. (\ref{A.7})  are field equations corresponding to the  Lagrangian
\beq{A.9}
 L_{TL} = \frac{1}{4}  \sum_{s,s' \in S} K^{-1}_s
   A_{s s'} \dot{q^s} \dot{q^{s'}}
  -  \sum_{s \in S} A_s  \exp( \sum_{s' \in S} A_{s s'} q^{s'} ).
\eeq

For the energy corresponding to the solution (\ref{A.5}) we get
\beq{A.10}
E=\frac12<\dot x,\dot x> + \sum_{s \in S} \exp(2<b_s,x>)
= E_{TL} + \frac12 <\alpha,\alpha>,
\eeq
where
\beq{A.11}
 E_{TL} = \frac{1}{4}  \sum_{s,s' \in S} K^{-1}_s
  A_{s s'} \dot{q^s} \dot{q^{s'}}
  + \sum_{s \in S} A_s  \exp( \sum_{s' \in S} A_{s s'} q^{s'} ),
\eeq
is the energy function  corresponding to the Lagrangian (\ref{A.10}).

For dual vectors $u^s\in V^*$ defined as
$u^s(x)=<b_s,x>$, $\forall x \in V$, we have $<u^s,u^l>_*=<b_s,b_l>$,
where $< \cdot, \cdot>_*$ is dual form on  $V^*$.  The orthogonality
conditions (\ref{A.6}) read
\beq{A.18}
u^s(\alpha)=u^s(\beta)=0 ,
\eeq
$s \in S$.

\section{Conclusions}

Here we obtained a family of solutions  in  multidimensional gravity
with  $p$-branes generalizing Majumdar-Papapetrou type solutions
>from \cite{IMC, IMBl} in a special case of one harmonic
function. These solutions are
related to Toda-like systems (of general type)
and are defined up to the
solutions of Laplace and Toda-type equations.
We  considered special solutions  corresponding to $A_m$ Toda lattices
(written in a parametrization of A. Andersen \cite{And}).
The general solutions may be also used for other open Toda lattices,
e.g. corresponding to $B_m$, $C_m$, $D_m$ series.
The solutions also contain a class of ``cosmological''
and spherically symmetric solutions and may be used for
investigation of possible black hole and wormhole configurations.

\begin{center}
{\bf Acknowledgments}
\end{center}

This work was supported in part
by the Russian Ministry for
Science and Technology, Russian Foundation for Basic Research,
project SEE, Korea Research Foundation made in the program year of 1997,
KOSEF No. 95-0702-04-01-3 and project SEE.
V.D.I also thanks  the members of
Department of Science Education and Basic Science Research Institute
of Ewha Womans University (Seoul, Korea) for the kind hospitality
during his visit in April-May of 1999.


\small
\begin{thebibliography}{99}

\bibitem{M-th1}
E. Witten, {\it Nucl. Phys.} {\bf B 443}, 85 (1995); hep-th/9503124; \\
P. Townsend, {\it Phys. Lett. } {\bf B 350}, 184 (1995); hep-th/9612121; \\
C. Hull and P. Townsend, {\it Nucl. Phys.} {\bf B 438}, 109 (1995);
hep-th/9610167; \\
P. Horava  and E. Witten, {\it Nucl. Phys.} {\bf B 460}, 506 (1996);
hep-th/9510209.

\bibitem{M-th2}
J.M. Schwarz,  Lectures on Superstring and M-theory Dualities,
hep-th/9607201; \\
M.J. Duff,  M-theory (the Theory Formerly Known as Strings),
hep-th/9608117.


\bibitem{GSW}
M.B. Green, J.H. Schwarz and E. Witten, Superstring Theory,
vol. 1, 2, Cambridge, 1987.

\bibitem{St}
K.S. Stelle, Lectures on Supergravity p-branes, hep-th/9701088.

\bibitem{DKL}
M.J. Duff, R.R. Khuri and J.X. Lu,
{\it Phys. Rep.} {\bf 259},  213 (1995).

\bibitem{Str}
A. Strominger, {\it Phys. Lett. } {\bf B 383}, 44 (1996);
hep-th/9512059.


\bibitem{CT}
M. Cvetic and A.A. Tseytlin,  Nucl. Phys. B 478, 181 (1996).

\bibitem{Ts1}
A.A. Tseytlin, {\it Nucl. Phys.} {\bf B 487}, 141 (1997);
hep-th/9609212.

\bibitem{PT}
G. Papadopoulos and P.K. Townsend,
{\it  Phys. Lett.} {\bf B 380}, 273 (1996); hep-th/9603087.

\bibitem{Ts}
A.A. Tseytlin,
{\it Nucl. Phys.} {\bf B 475}, 149 (1996); hep-th/9604035.

\bibitem{GKT}
J.P. Gauntlett,  D.A. Kastor, and J. Traschen,
{\it Nucl. Phys.} {\bf B 478}, 544 (1996); hep-th/9604179.

\bibitem{LPX}
H. L\"u, C.N. Pope, and K.W. Xu, Liouville and Toda Solitons in
M-theory, hep-th/9604058.

\bibitem{LMPX}
H. L\"u, S. Mukherji, C.N. Pope and K.-W. Xu,
Cosmological Solutions in String Theories,
hep-th/9610107.

\bibitem{V}
A. Volovich,
{\it  Nucl. Phys. }  {\bf B 487} (11), 141 (1997); hep-th/9608095.

\bibitem{AV}
I.Ya. Aref'eva and A.I. Volovich,
{\it Class. Quantum Grav.} {\bf B 14}, 29901 (1997);
hep-th/9611026.

\bibitem{IM0}
V.D. Ivashchuk and V.N. Melnikov,
Intersecting p-brane Solutions in Multidimensional
Gravity and M-theory, hep-th/9612089;
{\it Grav. and Cosmol.} {\bf 2}, No 4, 204 (1996).

\bibitem{IM}
V.D. Ivashchuk and V.N. Melnikov,
{\it Phys. Lett. } {\bf B 403}, 23 (1997).

\bibitem{BREJS}
E. Bergshoeff, M. de Roo, E. Eyras, B. Janssen and
J.P. van der Schaar, {\it Class. Quantum Grav.} {\bf 14} , 2757 (1997);
hep-th/9612095.

\bibitem{AR}
I.Ya. Aref'eva and O.A. Rytchkov,
Incidence Matrix Description of Intersecting p-brane
Solutions, {\it Preprint} SMI-25-96, hep-th/9612236.

\bibitem{AEH}
R. Argurio, F. Englert and L. Hourant,
{\it Phys. Lett. } {B 398}, 2991 (1997); hep-th/9701042.

\bibitem{AIR}
I.Ya. Aref'eva, M.G. Ivanov and O.A. Rytchkov,
Properties of Intersecting p-branes in Various Dimensions,
{\it Preprint} SMI-05-97, hep-th/9702077.

\bibitem{AIV}
I.Ya. Aref'eva, M.G. Ivanov and I.V. Volovich,
Non-Extremal Intersecting p-Branes in Various Dimensions, hep-th/9702079;
{\it Phys. Lett. } {\bf B 406}, 44 (1997).

\bibitem{Oh}
N. Ohta, Intersection Rules for Non-extreme p-branes, hep-th/9702164.

\bibitem{KKLP}
N. Khvengia, Z. Khvengia, H. L\"u and  C.N. Pope,
{\it Class. Quantum Grav.} {\bf 15}, 759 (1998);
hep-th/9703012.


\bibitem{IMC}
V.D. Ivashchuk and V.N. Melnikov,
Sigma-model for the Generalized  Composite p-branes,
hep-th/9705036; {\it Class. Quantum Grav.} {\bf 14}, 3001 (1997);
Corrigenda {\it ibid.} {\bf 15 } (12), 3941 (1998).

\bibitem{IMR}
V.D. Ivashchuk, V.N. Melnikov and M. Rainer,
Multidimensional $\sigma$-models with Composite Electric $p$-branes,
gr-qc/9705005; {\it Grav. and Cosmol. } {\bf 4}, No 1 (13), (1998).


\bibitem{BGIM}
K.A. Bronnikov, M.A. Grebeniuk, V.D. Ivashchuk and V.N. Melnikov,
Integrable Multidimensional Cosmology for
Intersecting $p$-branes,
{\it Grav. and Cosmol. } {\bf  3}, No 2(10), 105 (1997).

\bibitem{LMMP}
H. L\"u, J. Maharana, S. Mukherji  and C.N. Pope,
Cosmological Solutions, p-branes and the Wheeler De Witt
Equation,   hep-th/9707182.


\bibitem{GrIM}
M.A. Grebeniuk, V.D. Ivashchuk and V.N. Melnikov,
Integrable Multidimensional Quantum Cosmology  for Intersecting p-Branes,
{\it Grav. and Cosmol.\/} {\bf 3}, No 3 (11), 243 (1997), gr-qc/9708031.

\bibitem{BKR}
K.A. Bronnikov, U. Kasper and M. Rainer,
Intersecting Electric and Magnetic $p$-Branes: Spherically Symmetric
Solutions, gr-qc/9708058.



\bibitem{IMJ}
V.D. Ivashchuk  and  V.N. Melnikov, Multidimensional Classical
and Quantum Cosmology with Intersecting $p$-branes,
{\it J. Math. Phys.}, {\bf 39}, 2866 (1998); hep-th/9708157,

\bibitem{Y}
D. Youm, submitted to {\it Phys. Rep},  hep-th/9710046.


\bibitem{BIM}
K.A. Bronnikov, V.D. Ivashchuk and V.N. Melnikov,
The Reissner-Nordstr\"om Problem for
Intersecting Electric and Magnetic $p$-branes, gr-qc/9710054;
{\it Grav. and Cosmol.}, {\bf 3}, No 3(11), 203 (1997).


\bibitem{Br}
K.A. Bronnikov, Block-orthogonal Brane systems, Black
Holes and Wormholes, hep-th/9710207;
{\it Grav. and Cosmol.} {\bf 4}, No 1 (13),  49 (1998).


\bibitem{GR}
D.V. Gal'tsov and O.A. Rytchkov, Generating Branes via
Sigma models, hep-th/9801180.

\bibitem{IMBl}
V.D. Ivashchuk and V.N. Melnikov,
Madjumdar-Papapetrou Type Solutions in Sigma-model
and Intersecting p-branes,
{\it Class. Quantum Grav.} {\bf 16}, 849 (1999);
hep-th/9802121.

\bibitem{IKM}
V.D.Ivashchuk, S.-W.Kim and V.N.Melnikov, Hyperbolic Kac-Moody
Algebra from Intersecting $p$-branes, to appear in J. Math. Phys.;
hep-th/9803006.


\bibitem{GrI}
M.A. Grebeniuk and V.D. Ivashchuk,
Sigma-model Solutions and Intersecting p-branes
Related to Lie Algebras,
{\it Phys. Lett. } {\bf B 442}, 125 (1998);
hep-th/9805113.

\bibitem{GM1}
V.R. Gavrilov and V.N. Melnikov,
Toda Chains with Type $A_m$  Lie Algebra for Multidimensional Classical
Cosmology with Intersecting $p$-branes, In :  Proceedings of the
International seminar "Curent topics in mathematical cosmology", (Potsdam,
Germany , 30 March - 4 April 1998), Eds. M. Rainer and H.-J. Schmidt,
World Scientific, 1998,  p. 310; hep-th/9807004.


\bibitem{IMJ2}
V.D. Ivashchuk and V.N. Melnikov,
Multidimensional Cosmological and Spherically Symmetric Solutions
with Intersecting $p$-branes, gr-qc/9901001; \\
Cosmological and Spherically Symmetric Solutions
with Intersecting $p$-branes, to appear in J. Math. Phys.

\bibitem{CIM}
S. Cotsakis, V.D. Ivashchuk and V.N. Melnikov,
P-branes Black Holes and Post-Newtonian Approximation,
{\it Grav. and Cosmol.\/} {\bf 5}, No 1 (17), (1999); gr-qc/9902148.

\bibitem{GM2}
V.R. Gavrilov and V.N. Melnikov,
Toda Chains  Associated with   Lie Algebras  $A_m$ in Multidimensional
Gravitation and Cosmology with Intersecting $p$-branes,  submitted to
Theor. Math. Phys. (in Russian).

\bibitem{FS}
J. Fuchs and C. Schweigert, Symmetries, Lie algebras and
Representations. A graduate course for physicists
(Cambridge University Press, Cambridge, 1997).


\bibitem{CJS}
E. Cremmer, B. Julia, J. Scherk.
{\it Phys. Lett. } {\bf B 76}, 409 (1978).


\bibitem{MP}
S.D. Majumdar, {\it Phys. Rev. } {\bf 72}, 930 (1947);  \\
A. Papapetrou,  {\it Proc. R. Irish Acad. } {\bf A51}, 191 (1947).

\bibitem{NK}
G. Neugebauer and D. Kramer, {\it Ann. der Physik (Leipzig)}
{\bf 24}, 62 (1969).

\bibitem{KSMH}
D. Kramer, H. Stephani, M. MacCallum, and E. Herlt, {\it Exact
Solutions of the Einstein Field Equations\/}, CUP, 1980.

\bibitem{GC}
G. Cl\'ement, Gen. Rel. and Grav. {\bf 18}, 861 (1986);
{\it Phys. Lett.} {\bf A 118}, 11 (1986).


\bibitem{IMZ}
V.D. Ivashchuk, V.N. Melnikov and A.I. Zhuk,
{\it Nuovo Cimento } {\bf B 104}, 575  (1989).

\bibitem{T}
M. Toda, {\it Progr. Theor. Phys.} {\bf 45}, 174 (1970);
Theory of Nonlinear Lattices (Springer-Verlag, Berlin, 1981).




\bibitem{And}
A. Anderson, {\it J. Math. Phys.} {\bf 37}, 1349 (1996);
hep-th/9507092.

\bibitem{GIM}
V.R. Gavrilov, V.D. Ivashchuk and V.N. Melnikov,
{\it J. Math. Phys.} {\bf 36}, 5829 (1995).



\end{thebibliography}
\end{document}